\documentclass[twocolumn]{aastex631}

\usepackage{amsmath}
\usepackage{bm}
\usepackage{graphicx}
\usepackage{amssymb}
\usepackage{psfrag}

\def \d{\partial}
\newcommand{\dd}{\partial}
\def \bv{{\bf v}}

\def \br{{\bf r}}

\def \bB{{\bf B}}
\def \bp{{\bf p}}
\def \bR{{\bf R}}
\def \br{{\bf r}}

\def \brho{\boldsymbol{\rho}}

\def\bzeta{\boldsymbol{\zeta}}
\def \eps{\varepsilon}

\def\ls{\langle}
\def\rs{\rangle}

\submitjournal{ApJ}

\shorttitle{Non-Gaussian generalization of the Kazantsev-Kraichnan model}
\shortauthors{Kopyev et al.}
\graphicspath{{./}{figures/}}

\begin{document}

\title{Non-Gaussian generalization of the Kazantsev-Kraichnan model for turbulent dynamo}

\author{A.V. Kopyev}
\affiliation{P.N.Lebedev Physical Institute of RAS, 119991, Leninskij pr.53, Moscow, Russia}

\author{A.M. Kiselev}
\affiliation{P.N.Lebedev Physical Institute of RAS, 119991, Leninskij pr.53, Moscow, Russia}

\author{A.S. Il'yn}
\affiliation{P.N.Lebedev Physical Institute of RAS, 119991, Leninskij pr.53, Moscow, Russia}
\affiliation{National Research University Higher School of Economics, 101000,
Myasnitskaya 20, Moscow, Russia}

\author{V.A. Sirota}
\affiliation{P.N.Lebedev Physical Institute of RAS, 119991, Leninskij pr.53, Moscow, Russia}

\author{K.P. Zybin}
\affiliation{P.N.Lebedev Physical Institute of RAS, 119991, Leninskij pr.53, Moscow, Russia}
\affiliation{National Research University Higher School of Economics, 101000,
Myasnitskaya 20, Moscow, Russia}

\correspondingauthor{A.V. Kopyev}
\email{kopyev@lpi.ru}

\begin{abstract}
We consider a natural generalization of the Kazantsev-Kraichnan model for small-scale turbulent
dynamo. This generalization takes account of  statistical time asymmetry of a turbulent flow, and,
thus, allows to describe velocity fields with energy cascade. For three-dimensional velocity field,
generalized Kazantsev equation is derived, and evolution of the second order magnetic field
correlator is investigated for large but finite magnetic Prandtl numbers.
 It is shown that as $Pr_m \to \infty$, the growth increment tends to the limit known
 from the T-exponential (Lagrangian deformation) method.
Magnetic field generation is shown to be weaker than that in the Gaussian velocity field for any
direction of the energy cascade, and depends essentially on the Prandtl number.

\end{abstract}



\section{Introduction}
Magnetic field generation in turbulent plasma is one of the most probable mechanisms responsible
for  stellar, interstellar and intergalactic magnetism 
\citep[see, e.g.,][]{Zeld_book, Moffat, Parker, Brandenburg_rev, Astro-vse}. 
Small-scale turbulent dynamo has been the object of interest of many
researchers \citep[see, e.g.,][]{FGV,Brandenburg12, BiferaleRev} as it can provide intensive increase of magnetic
field \citep{Moffat, Kulsrud92}. In these problems, characteristic scale of magnetic field
fluctuations is much smaller than the scale at which turbulence is generated: this corresponds to
inertial and viscous scale ranges of turbulence.

The conception of small (seed) initial magnetic field fluctuations implies that there exists an
important stage of kinematic dynamo: magnetic field is small enough to cause no feedback on the
velocity distribution, so it is passively advected by the turbulent flow. Magnetic Prandtl number,
i.e. ratio of the kinematic viscosity~$\nu$ to the magnetic diffusivity~$\varkappa$, is the most important
characteristic of this advection process. In the paper we consider large Prandtl numbers:
$$
Pr_m = \nu / \varkappa \gg 1.
$$
Such situation is observed, e.g., in interstellar medium~\citep{Brandenburg_rev, Rincon}.
This means that the magnetic diffusive scale length~$r_d$ is much smaller than the Kolmogorov
viscous scale~$r_{\nu}$. We assume the characteristic scale length~$l$ of initial magnetic field
fluctuation to lie between these two scales,
$$
r_d \ll l \ll r_{\nu}.
$$

Evolution of magnetic field is described by a stochastic partial differential equation with random
velocity field acting as multiplicative noise.
 The velocity statistics is assumed to be stationary and known. The problem is to find the
 statistics of the magnetic field, in particular, its correlations.

 Kazantsev-Kraichnan model \citep{Kazantsev, KraichnanNagarajan} is the simplest and natural approximation for
 the velocity statistics: the velocity field is assumed to be Gaussian and $\delta$-correlated in
 time. In this model, all magnetic field  correlators are governed by the
 only two-point velocity correlator.

This model is an essential simplification.  
 Actually, unlike the additive random processes, in stochastic
equations with multiplicative noise 
the cumulants of all orders give comparable contributions to any statistical moment. So, the
Central limit theorem 'does not work' for these processes, and, to calculate even the second-order
correlators of magnetic field, one should use the Large deviation principle  and take all velocity
correlators into account. So, the replacement of arbitrary random velocity field by Gaussian
process can change the result crucially.

Besides, in Gaussian approximation for velocity field and hence, in Kazantsev-Kraichnan model,
there is no energy cascade.  %
 The energy of magnetic field excitation comes from the energy of the turbulent flow, which is
 generated at large scales; thus, the cascade may be important for dynamo.
 The non-zero third-order velocity correlator is responsible for energy
cascade and for time asymmetry in general \citep{K41, Frisch} : indeed, the inversion of
time would result in the change of sign of all velocities, and time symmetry implies that the
statistics would not change; hence, the third order correlator is zero for time symmetric flows.
Its presence indicates time asymmetry. So, the account of non-Gaussianity is highly desirable.

 There are two different theoretical approaches to investigate the magnetic field statistics.
 One of them is based on the Lagrangian deformations statistics \citep[see][]{ZMRS,Chertkov,epl}: it implies
 direct solution of the magnetic field evolution equation by means of the T-exponential
 formalism. The physical meaning of this method can be formulated in terms of
 independent magnetic blobs, each of them undergoing its evolution in  the turbulent
 flow \citep{MoffatSaffman,Kol2D,Scripta19,PhFlu21}. 
 This approach allows to calculate the magnetic field correlators of all
 orders, and to consider inhomogenous, in particular, localized initial magnetic field distributions.
 In this frame, it is possible to deal with arbitrary (not necessarily Gaussian) velocity
 statistics. So, this approach allows to consider velocity statistics
  wider than the Kazantsev-Kraichnan model.

  However, this approach is restricted to so-called Batchelor regime \citep{Batchelor}: the
  characteristic scale of the magnetic field must lie deep inside the viscous range of turbulence,
  so that the velocity field can be approximated by a linear function.  This means that the
  solutions found by this method are definitely applicable for some finite time range $ t \propto
  \ln r_{\nu}/l $.
  Later on, the characteristic scale of the magnetic field continues to increase and reaches the
  inertial range of turbulence. The 'Lagrangian deformation' approach may fail to predict the
behavior of correlators at this stage. The details of applicability of the method to the inertial stage
are considered by \citet{PhFlu21}.

The other approach  is based on statistical properties of pair correlators and allows to derive a
closed differential equation for the pair correlator of magnetic field (Kazantsev equation). It was
used and developed in many papers 
\citep[see, e.g.,][]{Kazantsev, Kraichnan, VainKitch, Kol2D, 2-5D, Sch-Cartesio,
Boldyrev_2007, KisIstomin13}; hereafter we will refer to it as to Kazantsev approach.
The advantage of the method is its applicability to any stage of the magnetic field evolution. However,
it is restricted to statistically homogenous magnetic field configurations, and it allows to
calculate the second-order two-point correlator only. There is one more vice of this approach: it
requires Gaussian and $\delta$-correlated in time velocity statistics, so it is restricted to the
Kazantsev-Kraichnan model of velocity field. Only for several models with some special additional
conditions its applicability has been enlarged \citep{tau-Sch, tau-Bhat14, tau-Rog}.

The two approaches produce concordant results wherever their domains of applicability overlap
\citep{Chertkov, PhFlu21}, they are also verified by numerical simulations
\citep{tau_DNS, Sch_highPm, recentDNS}.  
However, there remains the domain where neither of them can be applied: processes with non-Gaussian
and/or not $\delta$-correlated velocity statistics cannot be analyzed at late (inertial) stage of
their evolution neither by the Lagrangian deformations approach nor by the classical Kazantsev
method. The finite correlation time was taken into account in \cite{tau-Bhat14, tau_DNS, tau-Rog,
tau-Sch} for some special types of flows. The non-Gaussian velocity statistics in combination with
the inertial stage has not been considered yet.

To fill this gap, in this paper we consider the simplest non-Gaussian generalization of the
Kazantsev-Kraichnan model introduced in \cite{JOSS1, Scripta19}. It implies a non-zero third-order
velocity correlator, and thus, takes into account the time asymmetry of the flow. 
This model allows to investigate the long-time evolution of statistics for advection (and,
more generally, multiplicative) equations for arbitrary velocity field with small but non-zero
third order correlator.  
We generalize the Kazantsev method to apply it to this '$V^3$-model', and find the two-point pair magnetic field
correlator. We show that inside the Batchelor regime, the results obtained for the $V^3$-model by
means of the Lagrangian deformations approach and by means of the generalized Kazantsev method
coincide, which verifies the new generalized method. Now, for later (inertial) stage, we show that
the $V^3$ -model is stable relative to the limit of 'zero non-Gaussianity' as it turns into the
Kazantsev-Kraichnan model. We calculate the magnetic field growth increment for finite time
asymmetry, and evaluate the correction produced by finite Prandtl number.

It appears that small time asymmetry decreases the magnetic field generation, independently of the
direction of the energy cascade. The range of Prandtl numbers that produce effective generation is
also narrower as non-Gaussian time-asymmetry increases. The $V^3$-model is shown to be a useful and
effective instrument to investigate magnetic field advection in finite-Prandtl non-Gaussian flows.

The paper is organized as follows. In the next section, we 
briefly review the  basic ideas and equations of the two approaches 
by the example of the Kazantsev-Kraichnan model. Then we  recall the formulation and restrictions
of the $V^3$ model and the results obtained for this model by means of the Lagrangian deformations
method (Section III). In Section IV we derive the modified Kazantsev equation in order to apply the
Kazantsev approach to the  $V^3$ model. In the limit of infinite Prandtl number, it appears to be
possible to solve the equation and to find the increment of the pair magnetic field correlator. The
results of numerical solution of
the equation for finite Prandtl numbers, validation of the method 
and check of its stability is performed in Section V. In Discussion we analyze the obtained
results, and pay special attention to the applicability of $\delta$-correlated in time velocity
distribution and to comparison with the models with finite correlation time.

\section{Gaussian velocity field: recall of classic results}
To introduce the notations and equations needed, we start from the classical problem statement.
Kinematic transport of 
magnetic field $\bB(t,\br)$
advected by random statistically homogenous and isotropic nondivergent velocity field $\bv (t, \br)$, $\nabla \cdot \bv =0$,  is
described by the evolution equation
\begin{equation}  \label{1}
\frac{\partial \mathbf{B}}{\partial t}+\bigl(\mathbf{v} \nabla \bigr)\mathbf{%
B}-\bigl(\mathbf{B}\nabla\bigr)\mathbf{v} = \varkappa \Delta \mathbf{B},
\end{equation}
where $\varkappa$ is the diffusivity. The random process $\bv (\br,t)$ is assumed to be stationary,
and to have given statistical properties. The initial conditions for magnetic field are also
stochastically isotropic and homogenous. The aim is to find statistical characteristics of the
process $\bB$, in particular, its pair correlator.

From statistical homogeneity and isotropy, and non-divergency of $\bB$ it follows that its
simultaneous pair correlator has the form
\begin{multline}  \label{BBgeneral}
\left \langle B_i (\bR,t) B_j (\bR+\br,t) \right \rangle = G(r,t) \delta_{ij} + \\
\frac 12 r G'(r,t) (\delta_{ij}-r_i r_j/r^2).
\end{multline}
The average here is taken over the initial conditions $\bB(\br,0)$ and over the possible
 realizations of the velocity field $\bv(\br,t)$.

So, we are interested in time dependence of $G$: if it increases exponentially,
\begin{equation}  \label{gamma-def}
G \sim e^{\gamma t} G(r),
\end{equation}
 one calls the process 'turbulent dynamo', and $\gamma$ is called the magnetic field increment.

The Kazantsev-Kraichnan model implies that the velocity field is Gaussian, $\delta$-correlated in
time; its statistics is completely determined by the second-order correlator
\begin{equation}  \label{VV-general}
\left \langle v_i (\bR,t) v_j (\bR+\br,t+\tau) \right \rangle = D_{ij}(\br) \delta(\tau).
\end{equation}
The correlator does not depend on $\bR$ and $t$ because of homogeneity and stationarity. To make
contact with finite correlation time real flows, one can define $D_{ij}$ by
\begin{equation}  \label{VV-via-integral}
D_{ij}(\br) = \int \left \langle v_i (\bR,t) v_j (\bR+\br,t+\tau) \right \rangle d \tau.
\end{equation}
Just as in (\ref{BBgeneral}), non-divergency and isotropy oblige the tensor $D_{ij}$ to be
determined by only one scalar function of a scalar argument.
 For the purposes of next subsection, it is convenient to consider the
(time-integrated) longitudinal structure function
\begin{equation}  \label{structure function}
\begin{array}{r}
\sigma(r) = \frac 14 \int d\tau \left \langle  \left(  \left( \bv (\br,\tau) - \bv (0,\tau) \right)
{\br}/{r}
 \right) \right. \\
\left. \times  \left( \left( \bv (\br,0) - \bv (0,0) \right) 
{\br}/{r} \right)
\right \rangle.
\end{array}
\end{equation}
Then
$$
\sigma(r) = \frac 12 \left( D_{ij} (0) \frac{r_i r_j}{r^2}  - D_{ij} (r) \frac{r_i r_j}{r^2}.
\right)
$$
%
If one presents $D_{ij}$ in the form analogous to (\ref{BBgeneral}),
\begin{equation} \label{DcherezP}
D_{ij} =  P(r) \delta_{ij} + \frac 12 r P'(r) (\delta_{ij}-r_i r_j/r^2),
\end{equation}
then
$$
\sigma(r) = \frac 12 (P(0)-P(r)) \ .
$$
In presence of viscosity, the velocity field is smooth at the smallest scales, and
\begin{equation}  \label{defD}
  P(r) =_{r\to 0} P(0) - \frac 23 D r^2 + O(r^3) \ ,   D=-\frac 34 P''(0),
\end{equation}

\subsection{Kazantsev Equation}
The equation to describe the evolution of the second-order correlator (\ref{BBgeneral}) can be
found from Eq.(\ref{1}) by means of multiplying and subsequent averaging.  The cross-correlations
of magnetic field and velocity can be split by means of the Furutsu-Novikov theorem due to
Gaussianity and delta-correlation \citep{Furutsu,Novikov}:
\begin{equation}\label{FurNovikov_classic}
\left \langle {v_i(\br, t) g[\bv] } \right \rangle  = \frac 12 \int \, \mathrm{d} \br' D_{ij} ( \br
- \br') \left\langle \frac{\delta g[\bv]}{\delta v_j(\br',t)}  \right\rangle ,
\end{equation}
where $g[\bv]$ is an arbitrary analytic functional of $\bv(\br,t)$  
 and $\delta/\delta v$  is a functional derivative.
 Thus, for $G(\br,t)$ one gets
the equation
\begin{equation}  \label{Kaz-eq}
\frac {\partial G(\br,t)}{\partial t} = L_{Gauss} G \ ,
\end{equation}
$$
\begin{array}{rl} \displaystyle
L_{Gauss} =  & \displaystyle 2 \sigma(r) \frac{\partial^2}{\partial r^2} + 2 (\sigma'+\frac 4r
\sigma) \frac {\partial}{\partial r} + 2 (\sigma''+\frac 4r \sigma')
\\[0.2cm] & \displaystyle
 + 2 \varkappa \left(
\frac{\partial^2}{\partial r^2} +\frac 4r \frac{\partial}{\partial r} \right).
\end{array}
$$

In a turbulent hydrodynamic flow, $\sigma$ has the following asymptotics:
\begin{equation}  \label{sigma-ranges}
 \sigma (r) = \left\{ \begin{array}{ll} \frac{D}3 r^2 ,& \ r\ll r_{\nu}, \\
const \cdot r^{\xi} ,& \ r_{\nu}\ll r \ll L, \\
\frac 12 P(0) ,& \ r \gg L,
\end{array} \right.
\end{equation}
where $r_{\nu}$ is the viscous dissipation scale, and $L$ is the integral scale of turbulence.
 The time scale 
$$D^{-1} = \frac{2 r_\nu}{v_\nu}$$ 
is of the order of the eddy turnover time at the viscous scale \citep[see][]{Brandenburg_rev}. 
The first asymptote in (\ref{sigma-ranges}) corresponds to the viscous range of scales, the second presents
the inertial range, and the last string is for the integral range of turbulence \citep[see][]{LLVI}.

The well-known result \citep{tau-Rog,Sch-lnPm2002} is that for large Prandtl numbers, independently
of the parameter $\xi$ (characterizing the inertial range), the equation (\ref{Kaz-eq}) has a
growing mode with the increment
$$
\gamma = \frac 52 D - O(\ln^{-2} \frac{r_d}{r_{\nu}}),
$$
where
$$
r_d =  \sqrt{ \varkappa/D} \propto r_{\nu} / \sqrt{Pr_m}.
$$
Thus, turbulent dynamo exists at large Prandtl numbers, and the increment is determined
by the viscous range of turbulence.

\subsection{'Lagrangian Deformations' Approach}
This alternative way can only be applied for scales deep inside the viscous range, which
corresponds to the early stage of evolution of initially small-scale ($l\ll r_{\nu}$) magnetic
fluctuation \citep{Chertkov}. For scales much smaller than $r_{\nu}$, the velocity field is smooth.
So, one chooses a co-moving quasi-Lagrangian reference frame \citep{Lvov} associated to some fluid
particle $\br_0(t)$, and in this frame expands the (relative) velocity into a series up to the
first order:
$$
\delta v_i (\br,t) = A_{ij}(t) r_j.
$$
This is called Batchelor approximation \citep{Batchelor}, and $A_{ij}$ is the velocity gradient
tensor:
$$
A_{ij} = \partial_j v_i (\br_0(t),t).
$$
The transport equation (\ref{1}) now takes the form:
\begin{equation} \label{1A}
{\d_t} B_i + A_{jk} r_k 
\d_j B_i -A_{ij} B_j = \varkappa 
\Delta B_i
\end{equation}

In Kazantsev-Kraichnan model, the statistics of $A$ is determined by its second-order correlator:
$$
\left \langle A_{ij}(t) A_{kp}(t') \right \rangle =  D_{ijkp} \delta(t-t'),
$$
where, in accordance with (\ref{VV-general}),
$$
D_{ijkp}=-\partial_j \partial_p  D_{ik}({\bf 0}).
$$
The equation (\ref{1A}) can be solved explicitly by means of the Fourier transform \citep{ZMRS}; for
brevity, here we restrict ourselves to one spatial point and hence, to one-point correlator:
\begin{equation} \label{B-t}
B_m (t) = Q^{-1}_{mn} \int  B_n(\bp,0) e^{-\varkappa p_{i}p_j \int \left( QQ^T \right)_{ij}(t')
dt'} \mbox{d} {\bf p}.
\end{equation}
Here $Q(t)$ is the evolution matrix defined by the equation
\begin{equation} \label{Qdef}
\frac{dQ}{dt} = - QA, \quad Q(0)=1.
\end{equation}
It is convenient to use the polar decomposition for the evolution matrix:
$$
Q = s d R, \qquad s,R \in SO(3) , \qquad  d = \mbox{diag} \{ e^{-\zeta_i t} \} \ ,
$$
From incompressibility it follows $\det Q =1$, hence, $\zeta_1+\zeta_2+\zeta_3=0$.

It is well known
\citep{Furstenberg,Let} that the long-time asymptotic behavior of these three components is quite
different: as $Q$ obeys Eq. (\ref{Qdef}), $s(t)$ stabilizes at some random value that depends on
the realization of the process; $\zeta_i(t)$ are asymptotically stationary random processes and
tend (with unitary probability) to the limits $\lambda_i$,
$$
\lambda_1 \ge \lambda_2 \ge \lambda_3 \ ,
$$
the set of $\lambda_i$ is called the Lyapunov spectrum \citep{Oseledets}; and $R(t)$ rotates
randomly. We note that since $QQ^T = s d^2 s^T$ and $(Q^{-1})^T Q^{-1} = s d^{-2}
 s^T$, the matrix $R$ vanishes in the expression for $B^2(t)$. This simplifies the
calculation of the statistical moments. The Kazantsev -Kraichnan model provides one more
significant simplification: in particular, it corresponds to time-reversible flow, which means
$\lambda_2=0$.

Without loss of generality, the initial conditions for $B(\bp,0)$ can be chosen 
Gaussian, with pair  correlator
\begin{equation} \label{initialhom}
 \langle B_n(0,\bp) B_m (0,\bp') \rangle =  \delta(\bp+\bp')
  e^{-p^2 l^2} \left( p^2 \delta_{mn}  - {p_m p_n} \right),
 \end{equation}
Raising (\ref{B-t}) to the square and taking average over the initial conditions, one can obtain
\citep{Chertkov,ZMRS}
$$
\left \langle B^2 \right \rangle_{ic} \sim \left\{ \begin{array}{ll} d_2/d_1 ,& \ \ln d_2>0, \\
d_3/d_2 ,& \ \ln d_2<0.
\end{array} \right.
$$
To average this expression over all possible realizations of $A$, one considers the probability
density of $\zeta$:
\begin{align}
 \nonumber
&P({\bf y},t) = 
\langle \prod \limits_{j=1..3} \delta(\zeta_j(t)-y_j) \rangle \ , & \\
 \label{ukaz}
& \left \langle B^2 \right \rangle _{i.c., \bv}  = \int P(\bzeta,t)  \langle B^2 \rangle _{i.c.}
d\zeta_1 \dots d\zeta_3. 
\end{align}
The incompressibility condition leaves only two independent variables (e.g., $\zeta_1$,$\zeta_2$).
The probability density of $\zeta_j$ for any (not necessarily Gaussian) $A(t)$ can be expressed in
terms of statistics of the process $A(t)$ \citep{JOSS1,Scripta19}.

Eventually, for the Kazantsev-Kraichnan model one gets \citep{Chertkov}
\begin{align}
\label{KK-Batchelor}
&\lim \limits_{t\to \infty} \frac 1t \ln \left \langle B^2 \right \rangle  = \frac{5}{2} D, \\
&\lim \limits_{t\to \infty} \frac 1t \ln \left \langle B^{2n} \right \rangle  = 
\left( 2n+ \frac{n^2}2 \right) D.
\end{align}

We see that (\ref{KK-Batchelor}) coincides with the increment obtained in the Kazantsev approach.
So, it appears that the asymptote found in Batchelor approximation remains to be valid not only
during the initial stage, $t < \frac 1D \ln (r_{\nu}/l)$,  but also at later stages of evolution.
This fact is non-trivial, since, e.g., in two-dimensional flows the Kazantsev equations shows no
growing modes, and the exponential increase of magnetic field at the initial Batchelor stage
changes to decrease at larger time \citep{Sch-Cartesio,Kol2D}.

\section{$V^3$ model in the method of Lagrangian deformations}

To generalize the Kazantsev-Kraichnan model, one has to add higher-order connected correlators; in
particular, to take time asymmetric processes into account, one has to deal with third-order
correlators. The isotropy and incompressibility conditions reduce the degrees of freedom of the
whole tensor $\langle A_{ij} A_{km} A_{nl} \rangle $ to one arbitrary multiplier $F$. The general
expression for all the components is given by \cite{Pumir}; here we restrict our consideration to
the correlators of the diagonal elements $A_{jj}$ (no summation), since these components are the
only ones needed for calculation of $\langle B^2 \rangle$ \citep{JOSS1,Scripta19}:
\begin{equation} \label{AppAqq}
\langle A_{pp}(t)  A_{qq}(t') A_{rr}(t'') \rangle =F f_{pqr} \delta(t-t') \delta(t-t'') \ ,
\end{equation}
\begin{align*}
& f_{111}= f_{222}= f_{333}= f_{123}= - \frac 43,  \\ 
&f_{112}=f_{113}=f_{221}=...=\frac 23.
\end{align*}
We note that here $f_{pqr}$ is not a tensor.
 The right-hand side of (\ref{AppAqq}) is written in the form corresponding to an effective $\delta$-process \citep{JOSS1}.
The validlity of this approximation is verified by the possibility to reduce any finite-correlation
time non-Gaussian process to some delta-correlated process,  see Appendix A.

In the frame of the $V^3$ model, we set all the higher order connected correlators  zero.
A vice of this simplification is that the probability density is negative in some range
of its argument \citep{MY,Rytov} 
as only the second and the third connected correlators are unequal to zero. This
artefact can be fixed in the case of small $F$ by addition negligibly small but non-zero
higher-order correlators. These higher-order corrections would not affect the magnetic field
increment.

So, the time asymmetry of the velocity field in Batchelor regime is governed by only one parameter
$F$. The coefficient $F$ is the index of asymmetry of the flow. The  direction of cascade observed
in real three-dimensional hydrodynamic flows corresponds to $F>0$ \citep{reversible-lambda2}. 
The numerical simulation \citep{GirimajiPope} and the experiment \citep{Luthi} give an estimate of the
relation between the Lyapunov exponents $\lambda_2/\lambda_1\simeq 0.14$. 
The Lyapunov exponents are related to the
parameters $F$ and $D$ \citep{JOSS1,BF} by $ \lambda_1 / \lambda_2 = (2 D - F) / (2F) $.  %
So, this result for the Lyapunov spectrum corresponds to
 \begin{equation}  \label{0-13}
 F/D \simeq 0.13
 \end{equation}

\begin{figure}[t]   
\centering
\psfrag{f}{$F/D$}
\psfrag{m}{$\gamma/D$} 
\psfrag{l1}{(\ref{E:gamma_full})}  
\psfrag{l2}{(\ref{quasilagr_answer})}
\includegraphics[width=8cm]{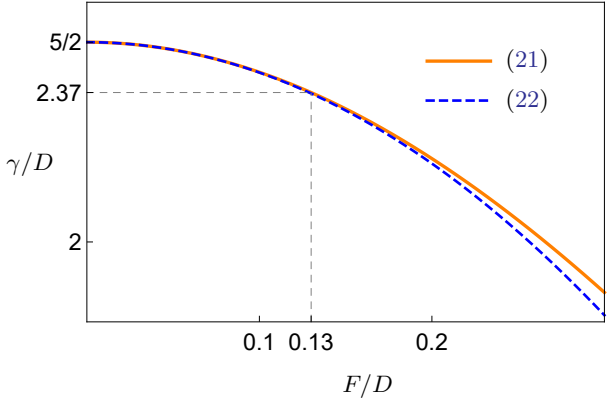}
\caption{$\gamma$ as a function of $F$, all values normalized by~$D$. The solid line corresponds to
(\ref{E:gamma_full}), and the dashed line represents (\ref{quasilagr_answer}).
 }\label{fig1}
\end{figure}

Returning to the calculation of $\langle B^2 \rangle$,
one makes use of the statistics of $A$ to calculate the probability density $P(\bzeta,t)$
\citep{JOSS1,Scripta19}; the long-time asymptote of the integral
in (\ref{ukaz}) can be found by the saddle point method, and the result is \citep{Scripta19}: 
\footnote{There is a misprint in the last formula of Section 10 in \cite{Scripta19};
here we give the corrected expression.}
\begin{equation}\label{E:gamma_full}
  \lim \limits_{t\to \infty} \frac 1t \ln \langle B^2(t) \rangle= \frac 49 \frac{D^3}{F^2}
+7D - \frac{\left(4D^2+27F^2\right)^{3/2}}{18F^2} \ ,
\end{equation}
In the limit  $F\ll 1$ we arrive to
\begin{equation}  \label{quasilagr_answer}
  \gamma= \frac{5}{2} D \left(  1-
\frac{243}{80} \left( \frac{F}{D} \right)^2 + o\left(\frac{F}{D}\right)^2 \right)
\end{equation}
We see that the average coincides with that for Kazantsev-Kraichnan model if $F=0$. Analogous
calculations for the higher order increments lead to
$$
  \lim \limits_{t\to \infty} \frac 1t \ln \langle B^{2n}(t) \rangle \simeq
    \left( 2n+\frac{n^2}2 \right) D
  - \frac 3{32} (2+n)^4 \frac{F^2}D
$$
The dependence $\gamma (F)$ for the exact and approximated equations (\ref{E:gamma_full}) and
(\ref{quasilagr_answer}) is presented in Figure~\ref{fig1}. One can see that the approximation
(\ref{quasilagr_answer}) works well for $F/D \lesssim 0.1$.

The expression (\ref{quasilagr_answer}) generalizes (\ref{KK-Batchelor}) for time asymmetric flows,
but it is still derived  for linear velocity field, and thus, is only valid at Batchelor stage of
magnetic field evolution. To investigate the dynamo generation at longer time, one should apply the
Kazantsev approach to the $V^3$ model.

\section{$V^3$ model in the generalized Kazantsev theory}

In the Kazantsev-Kraichnan model there is only one nontrivial velocity correlator:
\begin{equation} \label{vv=D}
\langle v_i (\br, t)  v_j (\br', t') \rangle =
D_{ij} ( \br - \br')\delta(t-t').
\end{equation}
 In the frame of
the $V^3$ model ideology, we add the third order correlator:   
\footnote{In fact, this expression is not accurate: to calculate the averages to get
the generalized Kazantsev equation, one needs to 'regularize' the $\delta$-functions and take the
limit of zero correlation time in the end of the calculation. So, in more accurate writing, the
arguments of the $\delta$-functions must be symmetrized. See Appendix B for more details.
\\
In Eq.(\ref{AppAqq}) the $\delta$-functions can also be symmetrized, but this is not necessary because
the Lagrangian deformations does not require the regularization.
}
\begin{equation} \label{vvv=F}
\langle v_i (\br, t)  v_j (\br', t')  v_k (\br'', t'') \rangle =
F_{ijk} (\br - \br', \br - \br'' ) \delta(t-t') \delta(t-t'').
\end{equation}

The correspondence with velocity gradients statistics (\ref{defD}),(\ref{AppAqq}) requires
\begin{align}
&D= - \frac 34 \frac {d^2}{dr^2} \left. \left( \frac 1{r^2} r_i r_j D_{ij} \right) \right| _{\br =0}, \\
&F= - \frac 34 \frac{\d^3}{\d r_1 \, \d r'_1 \, \d r''_1} F_{111}({\bf 0,0}).
\end{align}
Again, just as in the case of velocity gradients, the requirement of isotropy and incompressibility
of the flow leaves only one free parameter for the tensor $F_{ijk}$: it is the multiplier $F$ that
plays the role.

To apply the Kazantsev method to the non-Gaussian velocity field, one has to  use the non-Gaussian
version of the Furutsu-Novikov relation \citep{Klyackin, Rytov} to take the nonzero third order
correlator into account:
 \begin{equation}\label{FurNovikov_3}
\begin{array}{rl}
\langle v_i(\br, t) g[\bv] \rangle &= \frac 12 \int  dr' \, D_{ij} ( \br - \br') \left \langle
\frac{\delta}{\delta v_j(\br',t)} g[\bv] \right \rangle  \\
 +  \frac 16 \int  d\br'& d\br'' \,  F_{ijk} (\br - \br', \br - \br'') \left \langle
\frac{\delta^2}{\delta v_j(\br',t) \delta v_k(\br'',t)  } g[\bv] \right \rangle.
\end{array}
\end{equation}
In this equation the time integral is already calculated; see the details in Appendix B.

 Taking average of the square of Eq.(\ref{1}), we then arrive to the
modification of the Kazantsev equation (\ref{Kaz-eq})
\begin{equation}\label{dQdt_v3-1}
 \frac{\d G(r,t)}{\d t} = (L_{Gauss} +  \delta L) G(r,t).
\end{equation}
The expression for $\delta L$ is very cumbersome. Shorter expressions for important particular
choices of $F_{ijk}(r)$ will be presented in the next Subsections.

 We are interested in the long-time asymptotics 
 of the magnetic field correlator, so we seek for the solutions
$$
G(r,t) = e^{\gamma t} G(r) \ ,
$$
which transform (\ref{dQdt_v3-1}) into the ordinary differential equation
\begin{equation} \label{gammaG=}
\gamma G(r) = (L_{Gauss} +  \delta L) G(r).
\end{equation}

The important feature is that Eqs.(\ref{dQdt_v3-1}),(\ref{gammaG=}) are third order differential
equations with a small multiplier at the elder derivative. This results in appearance of a
non-physical solution that does not coincide with the solutions of (\ref{Kaz-eq}) as $F\to 0$;
instead, it goes to infinity. This solution must not be taken into consideration
 (see also more details in Appendix C).
 Technically, this non-physical solution is a consequence of
 the truncated sequence of correlators; if one adds higher order correlators of $A$ into
 consideration, the
 number of solutions of the Kazantsev equation would increase in accordance with the order of the
highest correlator. The  non-physical solutions would grow unrestrictedly as the magnitudes of the
higher order correlators tend to zero. These solutions have to be excluded by accurate choice of
the boundary conditions.

\subsection{Batchelor Regime}

In the Batchelor approximation, velocity gradients are assumed to be constant in space (although
dependent on time). Accordingly, the second derivatives of $D_{ij}$ and the third derivatives of
$F_{ijk}$ are assumed to be constant all over the liquid volume. Then, in accordance with
(\ref{DcherezP}),(\ref{defD}),
\begin{equation}  \label{sigma-ranges-1}
 \sigma (r) =  \frac{D}3 r^2.
\end{equation}
The exact expression for $F_{ijk}(\br-\br', \br-\br'') $ in the case is presented in Appendix B.
 Substituting this in (\ref{FurNovikov_3}), we get
\begin{equation} \label{27-5}
L_{Gauss}^B = \frac 23 D \left( r^2 \frac{\d^2}{\d r^2}  +6r \frac{\d}{\d r} +10 \right) + 
2 \varkappa \left( \frac{\d^2}{\d r^2} + \frac 4r \frac{\d}{\d r} 
\right),
\end{equation}
and the additional term in (\ref{dQdt_v3-1}),(\ref{gammaG=}) (see details of the derivation in Appendix B):
\begin{equation}   \label{deltaL-simple}
\delta L^B =\frac{1}{9} F \left(2r^3 \frac{\d^3}{\d r^3} + 21r^2 \frac{\d^2}{\d r^2} + 14r
\frac{\d}{\d r} - 70 \right). 
\end{equation}
The analytic analysis of Eq(\ref{dQdt_v3-1}),(\ref{sigma-ranges-1}),(\ref{deltaL-simple}) is
performed in Appendix C. We show that the fastest-growing mode corresponds to 
\begin{equation} \label{gammaFbatch}
\gamma = \frac{5}{2} D \left( 1- \frac{243}{80} \left( \frac{F}{D}\right)^2 +
o\left(\frac{F}{D}\right)^2 \right).
\end{equation}
This coincides with the exponents (\ref{quasilagr_answer}) found in the previous Section. The
important consequence from this expression is that, independently of the sign of $F$, the resulting
$\gamma$ is smaller than that found in the Kazantsev-Kraichnan model. This means that the magnetic
field generation is weaker in time-asymmetric flows than in the flow with Gaussian velocity
gradients, independently of the direction of the energy cascade. Also, $\gamma$ is monotonic
function of $F^2$: the time asymmetry of the flow decreases the generation.

The coincidence of the results obtained by means of   the Lagrangian deformation method and of the
modified Kazantsev approach is not trivial and not evident, and we will consider it in more details
in Discussion. Anyway, it proves that both methods work well as long as Batchelor approximation is
valid. However, the Kazantsev approach allows to investigate later stages of magnetic field
evolution, when characteristic scales of magnetic lines lengths exceed the viscous scale, and
spatial inhomogeneity of velocity gradients cannot be neglected.

\subsection{Nonlinear Velocity Field}
To take this inhomogeneity into account, one should consider scales comparable or larger than the
viscous scale:  outside this scale the correlators change essentially. In (\ref{sigma-ranges}) this
is expressed by means of three different ranges. We also have to cut off the third-order
correlator. Since it is known \citep{NovikovRuzmaikin,Kulsrud92} that (in the case of large $Pr_m$)
the details of the outer ranges do not affect the result significantly, we simplify
(\ref{sigma-ranges}) to
\begin{equation}  \label{sigma-ranges-late}
 \sigma (r) = \left\{ \begin{array}{ll} \frac{D}3 r^2 ,& \ r\le r_{\nu}, \\[0.1cm]
\frac{D}3 r_{\nu}^2 , & \ r > r_{\nu}.
\end{array} \right.
\end{equation}
In the $V^3$ model, we cut $F_{ijk}$ at the same boundary $r = r_{\nu}$.
 The first term in (\ref{gammaG=}) then takes the form%
\begin{equation} \label{Lgauss-nonlin}
\tilde{L}_{Gauss} = \begin{cases} L_{Gauss}^B  \ , \quad r \le r_{\nu} \ , \\
\left( \frac 23 D r_{\nu}^2  + 2 \varkappa \right) \left( \frac{\d^2}{\d r^2} +\frac{4}{r} \frac{\d}{\d r} \right), \quad r > r_{\nu} \ , \\
\end{cases}
\end{equation}
and the second term is
\begin{equation}  \label{deltaL-tilde}
\tilde{\delta L} = \begin{cases}  \delta L^B \ , r \le r_{\nu} \ , \\
0 \ , \ r>r_{\nu}.   \end{cases}
\end{equation}
Since we consider large magnetic Prandtl numbers, viscosity is large as compared to magnetic
diffusivity, and
\begin{equation}\label{def_rd}
r_d^2 = \frac{\varkappa}{D} \ll r_\nu^2.
\end{equation}
So, in addition to the small parameter $F/D$ in the Batchelor problem statement, here we have one
more small parameter
$$
1/\mu = r_d/r_{\nu}.
$$
 The equation (\ref{gammaG=}) with
(\ref{sigma-ranges-late}), (\ref{deltaL-tilde}) can be solved in special functions. However, we
make an analytic estimate for the contribution of large but finite $\mu$ to the magnetic increment:
\begin{equation} \label{itog}
   \gamma_{max}\gtrsim D \left( \frac{5}{2}-\frac{243F^2}{32D^2}-\frac{2\,\pi^2}{3\ln^2 \mu}\left(
   1+O(F^2/D^2)\right)
   \right)
\end{equation}
(See Appendix C for the derivation). This estimate shows that, in the first approximation, the
non-Gaussianity and the finiteness of the magnetic Prandtl number ($\mu < \infty$) act
independently.

 In the next Section we will  consider the numerical solutions to the
equation (\ref{gammaG=}) for finite $\mu$, investigate the dependence of magnetic field generation
on the parameters $F$ and $\mu$, and check the reliability of the model.

\section{Numerical solution of the generalized Kazantsev equation}
The generalized Kazantsev equation (\ref{dQdt_v3-1}) is a third-order differential equation with
small multiplier $F/D$ at the elder derivative. It is  not evident if its solutions are stable and
converge to the solutions of (\ref{Kaz-eq}) in the limit $F/D \to 0$.
We also test our theoretical conclusion (\ref{itog}) and show that the behavior of the increment
does not depend on the details of the cutoff at large $r$.

\subsection{Technical Details}
We consider the Eq.(\ref{gammaG=}) with $\sigma$ and $\delta L$ determined by
(\ref{sigma-ranges-late}) and (\ref{deltaL-tilde}). In dimensionless notations 
$$
x=r/r_d \  , \ \ \Gamma = \gamma /D \ , \  \ f=F/D \ , \ \ \mu = r_{\nu}/r_d
$$
we get the equation for $G(x)$:
\begin{equation} \label{dimensionless-eq}
\begin{array}{rl}
\Gamma G &= \frac 23 \left(  x^2 \theta(\mu -x) + \mu^2 \theta(x-\mu) +3 \right) G'' \\
&+ \frac 43 \left( 3x \theta(\mu -x) + 2 \frac{\mu^2}x \theta(x-\mu) +\frac 6x \right) G' \\
 &+ \frac {20}3 \theta(\mu-x)G \\
  &+ \frac 19 f \left( 2x^3 G'''+21 x^2 G'' + 14xG' - 70 \right) \theta(\mu-x),
  \end{array}
\end{equation}
where $\theta(y)$ denotes the Heaviside function.

The solution $G(x)$ depends on two parameters $f$ and~$\mu$. The asymptotes of the solutions
can be found analytically.

In the limit $x\to 0$, (\ref{dimensionless-eq}) has three modes:
\begin{align}
\label{left-asympt}
&G_{(1)} = 1 - Y x^2, \qquad  Y = \frac{1}{3} - \frac{7 f}{18} - \frac{\Gamma}{20}, \\
&G_{(2)} (x) \sim x^{-3}, \quad
G_{(3)} (x) \sim  x^6 \exp(\frac{9}{2 f x^2}).
\end{align}

The first two of them are close to the corresponding solutions for the Kazantsev-Kraichnan model;
the third one is 'produced' by the third-order term. The last two modes diverge as $x\to 0$, so,
they must be excluded from the physical solution.

The other asymptote is  $x\gg \mu$. The equation is significantly simplified in this limit because
$\sigma(r)$ becomes a constant:
\begin{equation}
p^2 G = G'' + \frac{4G'}{x} \ , \quad p^2 = \frac{\Gamma}{2 (1+ \frac{1}{3} \mu^2 )}.
\end{equation}
The exact solution to this equation is
\begin{equation}\label{right-asympt}
G(x) = Y_1 \frac{e^{-px}}{x} (1 + \frac{1}{px}) + Y_2 \frac{e^{px}}{x} (1 - \frac{1}{px}).
\end{equation}
Again, the divergency condition $G(x)\to_{x\to \infty} 0$ requires $Y_2=0$ and leaves only one of
the two modes.

To solve equation~(\ref{dimensionless-eq}) numerically, we fix some $\Gamma$ and the initial point $x_1\ll 1$. The initial conditions $G(x_1)$,$G'(x_1)$,$G''(x_1)$ are determined by the asymptote~(\ref{left-asympt}).  
Then we get the numerical solution $G(x)$ up to $x=\mu$, and match it with
the asymptote (\ref{right-asympt}); the condition $Y_2=0$  singles out a discrete spectrum of
possible values $\Gamma$.  We are looking for the maximal value $\Gamma = \Gamma_{max}$.
We also check the stability of the solution relative to the choice of $x_1$.

\begin{figure}[t]   
\centering \psfrag{f}{$f$}
\psfrag{g}{$\Gamma$} \psfrag{a1}{$\mu=30$} \psfrag{a2}{$\mu=100$}
\psfrag{a3}{$\mu=300$} \psfrag{t}{fit}
\includegraphics[width=8cm]{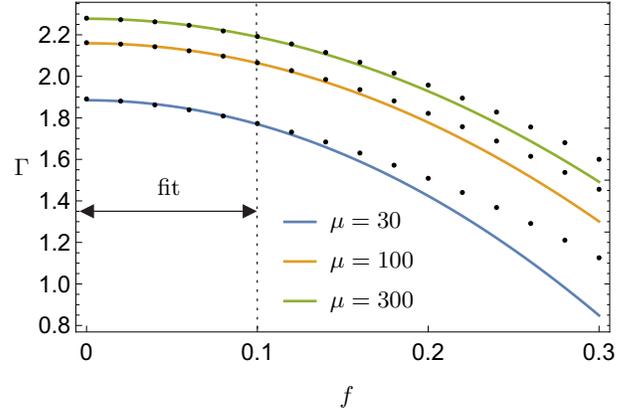}
\caption{Numerically calculated $\Gamma (f)$  for $\mu = 30$ (lower band), $100$ and $300$
(upper band)
fitted by   parabolic functions {\it inside the range } $0<f<0.1$. The declination at higher~$f$ is
a result of higher-order terms in (\ref{teor}).  }\label{fig2}
\end{figure}

\subsection{Results}

As it follows from (\ref{itog}), the theoretical prediction for $\gamma(f, \mu)$ is
\begin{equation} \label{teor}
\begin{array}{c}   \displaystyle
  \Gamma  = \frac{\gamma}{D} \simeq  
  \left( \frac 52 -c_2 f^2 -  \frac{c_3 +O(f^2) }{\ln^2 \mu} \right) \ , \\
c_2 = \frac{243}{32}\simeq 7.59 \ , \  c_3 \lesssim \frac{2\pi^2}3 \simeq 6.6
\end{array}
 \end{equation}

First, we analyze the dependence $\Gamma(f)$ for some fixed~$\mu$. The results for three values of
$\mu$ are presented in Figure~\ref{fig2}. We see that, in accordance with (\ref{teor}), $\Gamma(f)$ has
parabolic shape at small $f$ for all considered values of~$\mu$.
 To observe the dependence of the second-order term on~$\mu$, we fit the results of the simulation
 by  parabolic functions in the range $0\le f \le 0.1$;
 at larger~$f$, the higher-order terms in (\ref{teor}) may come into play.
 The results are presented in Table 1. One can see that the absolute term  is smaller than 2.5 and
 increases as a function~$\mu$. The magnitude of the declination agrees with (\ref{teor}).
 The absolute value of the coefficient at $f^2$ decreases as a function of~$\mu$, approaching the
 theoretical prediction $c_2$  for $\mu = \infty$.

\begin{table}[h]
\caption{Fit of the numerical data (Fig.2) for $0<f<0.1$:   $\Gamma(f)=C_1 - C_2 f^2$
\newline
}
\label{Table1}%
\begin{ruledtabular}
\begin{tabular}{cccc}
           $\mu$      & $C_1$ & $C_2$  \\
\hline
30 & $1.884 \pm 0.003 $& $11.5 \pm 0.5$ \\
100 & $2.159 \pm 0.001 $& $9.5 \pm 0.2 $\\
300  &  $2.277 \pm 0.001 $     &  $ 8.7 \pm 0.3 $      \\
\hline $\infty$ (theory) & 2.5  &  7.59 \\
\end{tabular}
\end{ruledtabular}
\end{table}

\begin{figure}[t]   
\centering
\psfrag{m}{$\mu$} \psfrag{g}{$\Gamma$} \psfrag{l1}{$f=0$} \psfrag{l2}{$f=0.13$}
\includegraphics[width=8cm]{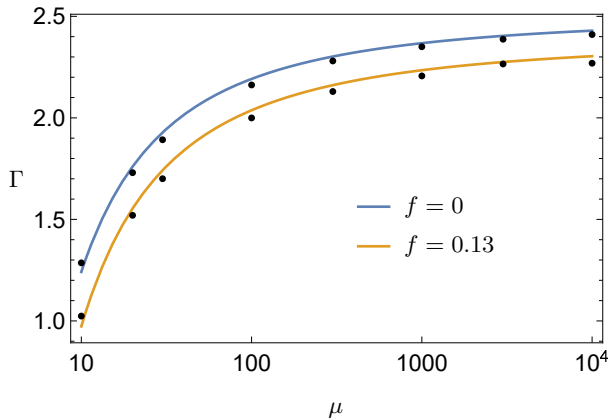}
\caption{ Numerical calculation of $ \Gamma (\mu)$  for $f=0$ (upper set of dots) and $f=0.13$
(lower set of dots) fitted by (\ref{gamma-mu-fit}). }\label{fig3}
\end{figure}

Second, we calculate the function $\Gamma(\mu)$ at some given~$f$. We take $f = 0.13$ because it
corresponds to the asymmetry of a real flow (\ref{0-13}) observed  in the numerical calculation
\citep{GirimajiPope} and experiment \citep{Luthi}. We also calculate $\Gamma(\mu)$ for a symmetric
flow, $f=0$: in this case, the equation (\ref{dimensionless-eq}) becomes a second order equation.
The results are presented in Figure~\ref{fig3}. We fit the graphs by the ansatz
\begin{equation} \label{gamma-mu-fit}
 \Gamma (\mu) = C_1-\frac{C_3}{\ln^2 \mu}.
\end{equation}
The correspondence is good enough; for $f=0$, we get  $C_1=2.50 \pm 0.02$ and  $C_3 = 6.64 \pm
0.33$, which coincides with the theoretical prediction  $c_1$ and $c_3$ in (\ref{teor}).

The choice of the $\sigma(r)$ profile at $r\simeq r_{\nu}$ is rather conditional; the details of
the velocity structure function at this range of scales are believed not to effect the result
significantly.  This has been checked numerically \citep{NovikovRuzmaikin} for  the
Kazantsev-Kraichnan model, but it has to be also proved for $F\ne 0$. 
So, apart from (\ref{sigma-ranges-late}), we consider a smooth  function
\begin{equation}\label{Sx_special}
\tilde{\sigma}(r)  = \frac{D}{3} \times \frac{(r/r_d)^2}{ 1 + (r/r_{\nu})^2 }\ .
\end{equation}
At $r\to \infty$ and $r\to 0$ it has the same asymptotes as $\sigma(r)$. We perform the same
calculations with this function. In Figure~\ref{fig4}, the dependence $\Gamma(\mu)$ for $f=0.13$ is presented
for both choices of $\sigma$.
We see that the details of the saturation do not affect the increment behavior crucially.

\begin{figure}[t]  
\centering
\psfrag{m}{$\mu$} \psfrag{g}{$\Gamma$} \psfrag{l1}{step} \psfrag{l2}{smooth}
\includegraphics[width=8cm]{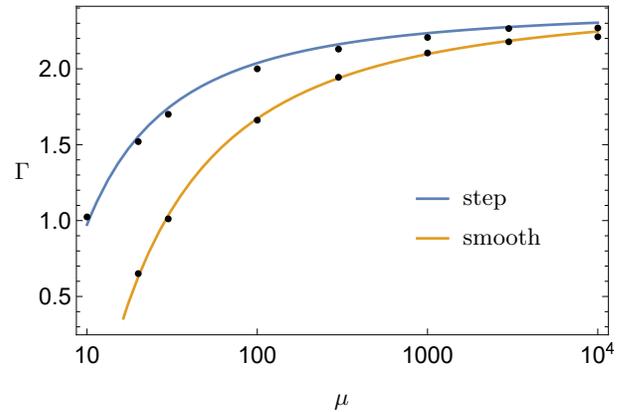} \label{fig4}  
\caption{ Dependence of the results on the shape of the velocity structure function.
  $\Gamma(\mu)$ for $f=0.13$
in the case of the step-like $\sigma(r)$ (\ref{sigma-ranges-late}) (upper dots) and the smoothed
$\tilde{\sigma}(r)$ (\ref{Sx_special})  (lower dots) fitted by (\ref{gamma-mu-fit}).}
\label{fig4}
\end{figure}

\begin{figure}[t]   
\centering
\psfrag{f}{$f$} \psfrag{m}{$\mu_{crit}$}
\includegraphics[width=8cm]{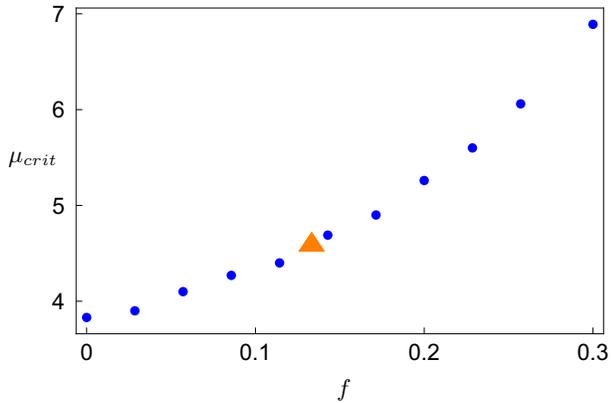}
\caption{The boundary of the magnetic field generation range $\gamma>0$, $\mu_{crit}$ as a function
of $f$. The triangle corresponds to the  point (\ref{0-13})}. \label{fig5}
\end{figure}

Finally, from (\ref{teor}) it follows that $\gamma(f,\mu)$ becomes negative outside some region in
the $f,\mu$ plane. So, the range of the parameters at which the generation of magnetic field is
possible is restricted by some $\mu > \mu_{crit} (f)$. The numerical calculations confirm this
prediction. We also
find the dependence $ \mu_{crit} (f)$: it is presented in Figure~\ref{fig5}. One can see that the presence of
non-Gaussian term decreases the range of $\mu$ that permits
 the generation.

\section{Discussion}

Thus, in this paper we consider the magnetic field generation in a turbulent hydrodynamic flow.
We derive the generalization of the Kazantsev equation for the case of time
asymmetric flows with small but finite third order velocity correlator, which is, probably, the
general case of real hydrodynamic turbulent flows.  The non-zero third order correlator corresponds to
the time asymmetry of the flow and is responsible for the energy cascade.

We use the $V^3$ model: 
slightly non-Gaussian stochastic velocity field is replaced by an effective $\delta$-process, and the velocity connected correlators of the order higher than three are assumed to be zero.
The validity of this model is argued in Section III and Appendix A.
We also use one more (unessential) simplifying assumption on the shape of velocity structure
function, which is considered piecewise.

The main results are: \\
\begin{itemize}
\item We show that the magnetic field generation weakens in presence of time asymmetry,
independently of the direction of the energy cascade (\ref{itog}): the increment of average
magnetic energy density decreases proportionally to the square of the magnitude $F$ of the third
order correlator.
\item   The range of magnetic
 Prandtl numbers (presented in the considered model by the parameter~$\mu$)
  that allow the magnetic field generation also becomes narrower for finite $F$: the critical
  Prandtl number increases as a function of $F$ (Figure~\ref{fig5}).
\item We show that, despite the presence of a small higher-order derivative, the numerical solution
of the generalized Kazantsev equation converges in the limit $F\to 0$ to the solution of the
Kazantsev equation. This proves that %
the generalized Kazantsev equation allows to investigate
numerically  the magnetic field evolution
  in time-asymmetric flows
   with intermediate   magnetic Prandtl numbers.
\end{itemize}

Now we proceed to the discussion of some interesting particular consequences.

 We solve the generalized Kazantsev equation analytically in the
limit of the Batchelor regime,
 $r_{\nu}=\infty$. In
this limit, the resulting increment (\ref{gammaFbatch}) coincides with the result that follows from
the Lagrangian deformation method (\ref{quasilagr_answer}). This coincidence
%
not only proves the reliability of both methods. It also reflects a non-trivial physical fact: the
time-averaged magnetic energy density measured  at some fixed point coincides with that measured
along a liquid particle trajectory.
 Not only the methods of calculation differ in the two approaches but also
 the averages $\langle B^2 \rangle$ are taken over
different ensembles.
 So, the coincidence not only verifies the results
but also proves the equivalence of these two averages. This coincidence is not a consequence of
ergodicity, since the trajectory of a particle as well as magnetic energy are functionals of the
velocity field, and, thus, are not independent.

The equivalence of the two average  was found for passive
scalar \citep{BF} and vector \citep{Chertkov} advection in the case of Kazanatsev-Kraichnan velocity
field. Now we see that it also holds for vector advection in  time-asymmetric velocity fields. 

Another important coincidence establishes the relation between the magnetic increments calculated
for the Batchelor limit and for the case of finite Prandtl number. Namely, the theoretical analysis
and numerical simulation confirm that, as $\mu \to \infty$,  the increment $\gamma (\mu, F)$
converges to the value $\gamma(F)$ found for the Batchelor regime. This is also not trivial: for
instance, this equality does not hold for two-dimensional flows \citep{Kol2D,Sch-Cartesio} or for
higher-order correlators in three-dimensional flows \citep{feedback}.
The Kazantsev equation corresponds to the infinite-time limit; the convergence of the solutions in the
limit $Pr_m \to \infty$ (and its coincidence with the value obtained for the Batchelor case) means
that the limits $Pr_m \to \infty$ and $t\to \infty$ commutate. This also has been known for
Gaussian flows \citep{Kulsrud92,Kazantsev,NovikovRuzmaikin}, now this is also checked for the time
asymmetric case.

 Eventually, the deformation of the magnetic energy spectrum and its evolution is also an
interesting and important question 
\citep{Kazantsev, Kulsrud92,  Sch-Cartesio, tau-Bhat14, a11}.
In the frame of $V^3$ model it is also possible to find the specrum evolution, however this problem is rather complicated and deserves separate investigation.
We will explore it in the next paper.

Summarizing, we stress that the asymmetry of the hydrodynamic flow statistics is an essential
feature that may affect the process of magnetic field generation.  The  $V^3$ model allows to
investigate  the asymmetric flows, and thus  opens the prospective for investigation of  transport
problems in the flows with energy cascade.

\newpage

\begin{acknowledgments}
The authors are grateful to Professor A.V.~Gurevich for his permanent attention to their work. 
A.M.K. thanks Professor Ya.N.~Istomin for considerable contribution to the work on its early stage.
The work of A.V.K. and A.M.K. was supported by the RSF Grant No. 20-12-00047.
\end{acknowledgments}

\appendix


\section{Grounds for the effective $\delta$-process introduction}

The equation 
(\ref{1A}) is a stochastic differential equation, the random velocity gradient tensor $\bf A$ acts as a
multiplicative noise \citep{Batchelor}.

One can show that for any non-Gaussian process with finite corelation time, there exists some corresponding 
effective $\delta$-process \citep{JOSS1}.
This means that 
\begin{equation}
\lim_{T \to \infty} \frac 1T \log  \langle B^2 (T) \rangle  = 
\lim_{T \to \infty} \frac 1T \log  \langle B^2 (T) \rangle_{\text{eff}} \, , 
\end{equation}
where the average in the right-hand side is calculated for velocity statistics defined by the effective $\delta$-process. 
The reason to replace an arbitrary finite-correlation time process by the corresponding $\delta$-process is that in the equation with multiplicative noise, the higher order connected correlators of the noise contribute to the long-time statistical properties of the solutions only via their integrals. 
This allows to substitute singular correlation functions for the real correlators.

We demonstrate this fact by a simple example.
Consider a one-dimensional stochastic equation with multiplicative noise $\xi(t)$:
\begin{equation} \label{AA1}
\partial _t x(t) = \xi (t) x(t) \ , \ \ x(0)=0,
\end{equation}
where $\xi(t)$ is a continous stationary random process with finite correlation time. Let it have
regular fast decaying connected correlation functions (cumulants):
$$
\langle \xi(t_1) \dots \xi(t_n) \rangle_c = W^{(n)} (t_1 - t_2, \dots, t_1 - t_n).
$$
The cumulant generating functional is defined by
\begin{equation} \label{AA2}
\langle e^{\int \xi(t) \eta(t) dt} \rangle = e^{W \left[ \eta(t) \right] } \ ,
\end{equation}
then
\begin{equation} \label{AA3}
W [ \eta(t) ] = \sum \limits _n \frac 1{n!} \int W^{(n)} (t_1 - t_2, \dots, t_1 - t_n) \eta(t_1)
\dots \eta(t_n) dt_1 \dots dt_n
\end{equation}

The solution of the Eq.(\ref{AA1}) for each continuous realization of $\xi(t)$ can be written as
\begin{equation} \label{AA4}
x(T) = e^{\int \limits _0 ^T \xi(t) dt}
\end{equation}
We are interested in the statistical moments
$$
\langle x^m (T) \rangle = \langle e^{m \int \limits _0 ^T \xi(t) dt} \rangle
$$
From (\ref{AA2}) it then follows that
$$
\langle x^m (T) \rangle = e^{W[m \theta(t) \theta(T-t)]},
$$
where $\theta$ is the Heaviside step-like function. According to (\ref{AA3}), we then have
$$
\lim \limits_{T \to \infty} \frac 1T \log  \langle x^m (T) \rangle  = w(m) \ ,
$$
$$
w(m) = \sum \limits _n \frac {m^n}{n!} w^{(n)}.
$$
$$
w^{(n)}  = \int W^{(n)} (\tau_2, \dots, \tau_n) d\tau_2 \dots d\tau_n,
$$
We see that as $T \to \infty$ the statistical moments of $x$ depend only on the integrals $w^{(n)}$ 
and do not depend on the detailed shape of the correlators. So, for any random process $\xi(t)$
(i.e., for any given $W^{(n)}$), we can consider a series of random processes $\xi_{\epsilon} (t)$
with
$$
W_{\epsilon}^{(n)} = \frac 1{\epsilon^{n-1}} W^{(n)} (\tau_2/\epsilon, ..., \tau_n/\epsilon)
$$
and all these processes would produce the same long-time asymptotes for the moments of $x(t)$. The
effective process $\xi_{eff}(t)$ is defined as the formal limit of this consequence as $\epsilon
\to 0$. Its connected correlation functions are
$$
\langle \xi_{eff}(t_1) \dots \xi_{eff}(t_n) \rangle_c = w^{(n)} \delta(t_1 - t_2) \dots \delta (t_1
- t_n)
$$
In the case of multidimensional stochastic processes and stochastic fields the expression for the
exact solution looks much more complicated than (\ref{AA4}), since the integral transforms into a
continuous matrix product (T-exponent). However the long-time
asymptotes of the solutions' moments still depend only on the integrals of the connected
correlators of the noise. This is the reason and the justification for substitution of the
effective $\delta$-process for {\bf \it any} random process. This tool is convenient in turbulence and
turbulent transport problems, since it allows to get closed equations for statistical moments.

\mdseries  

\section{Derivation of the generalized Kazantsev equation}

We start with Eq. (\ref{1}); to get the equation for the pair correlator, we multiply it by $\bf B$ and
take the average. For brevity, we denote
$$
\begin{array}{c}
B_{\alpha} = B_{\alpha}(\mathbf{r},t), \quad B'_{\alpha} = B_\alpha(\mathbf{r}',t), \\ 
v_\alpha=v_\alpha(\mathbf{r},t) , \quad  v'_\alpha=v_\alpha(\mathbf{r}',t) ,
\\
\d_\alpha = \frac{\d}{\d r_\alpha}, \quad\d'_\alpha = \frac{\d}{\d r'_\alpha}
,    \quad\d^{\rho}_\alpha= \frac{\d}{\d (\br'-\br)_\alpha}.
\end{array}
$$
We also take into account that homogeneity of the flow, which results in
$\d_m \langle ...\rangle = - \d'_m \langle ...\rangle =- \d^{\rho}_m \langle ...\rangle  $ and \\
$\ls v'_p B'_q B_\alpha \rs=-\ls v_p B_q B'_\alpha \rs$.
Then the equation for the pair correlator takes the form
\begin{equation}  \label{AppB27}    
\frac{\d}{\d t} \langle B_\alpha B'_\beta \rangle  =
- \varepsilon_{\alpha mn} \varepsilon_{n p q} \d_m^{\rho} \ls v_p B_q B'_\beta \rs
- \varepsilon_{\beta mn}
\varepsilon_{n p q}
\d_m^{\rho} \ls v_p B_q B'_\alpha \rs
+ 2\varkappa\, \d^{\rho}_m \d^{\rho}_m \ls B_\alpha B'_\beta \rs.
\end{equation}
Now we have to split the mixed correlations by means of the generalized Furutsu-Novikov equation
\citep{Furutsu,Klyackin}:
\begin{multline}\label{Fur-Nov-detailed}
\ls{v_p B_q B'_r}\rs  =\int \, d\mathbf{r}_1 dt_1 \ls v_p(\mathbf{r},t) v_{i_1}(\mathbf{r}_1,t_1)\rs\,
\left \langle \frac{\delta \bigl(B_q(\mathbf{r},t) B_r(\mathbf{r}',t)\bigr)}
{\delta v_{i_1}(\mathbf{r}_1,t_1)}\right \rangle
\\
+ \frac 1{2}  \int \, d\mathbf{r}_1 dt_1d\mathbf{r}_2 dt_2 \ls v_p(\mathbf{r},t)
v_{i_1}(\mathbf{r}_1,t_1) v_{i_2}(\mathbf{r}_2,t_2)\rs\,
\left \langle \frac{\delta^2 \bigl(B_q(\mathbf{r},t) B_r(\mathbf{r}',t)\bigr)}
{\delta v_{i_1}(\mathbf{r}_1,t_1)\delta v_{i_2}(\mathbf{r}_2,t_2)} \right \rangle.
\end{multline}
Since the variational derivatives contain delta-functions, to avoid the products  of
delta-functions with coinciding arguments,  we have to deal more accurately with the time
coincidence in correlators. Namely, we have to introduce a 'regularized $\delta$-function': a
bell-shaped function $\delta_{\epsilon}(\tau)$ satisfying the normalization condition $\int
\delta_{\epsilon}(\tau) d\tau = 1$ and with the width of the order of the correlation time.
Then (\ref{vv=D}) and (\ref{vvv=F}) can be written more precisely:
\begin{equation}
\ls{v_i(\mathbf{r},t) v_j(\mathbf{r}_1,t_1)}\rs = 
D_{ij}(\br - \br_1)  \delta_\eps(t-t_1),
\end{equation}
and
\begin{equation}
\ls{v_i(\mathbf{r},t) v_j(\mathbf{r}_1,t_1) v_k(\mathbf{r}_2,t_2)}\rs
= \frac{1}{3} F_{ijk}(\br-\br_1, \br-\br_2) \,\Bigl(\delta_{\eps}(t-t_1)\delta_{\eps}(t-t_2) +
\delta_{\eps}(t-t_1)\delta_{\eps}(t_1-t_2)
+\delta_{\eps}(t-t_2)\delta_{\eps}(t_1-t_2)\Bigr).
\end{equation}

The  isotropy and homogeneity conditions imply that the expression for $F_{ijk}$ can be written as
a composition of $\delta$-symbols and the components of $\bf a$, $\bf b$. The  non-divergency
condition results in the requirement $\d F_{ijk}/\d a_i = \d F_{ijk} /\d b_j = (\d/\d a_k + \d/ \d
b_k) F_{ijk}=0$; also, the existence of viscosity means that $F_{ijk}$ is proportional to $r^3$,
i.e., is a third-order polynom. These conditions determine the tensor  $F_{ijk}$ to a constant
multiplier:
\begin{multline}
F_{ijk}({\bf a},{\bf b}) = - \frac 1{18} F
\Bigl((2a^2-15b^2-2\,\mathbf{a}\cdot\mathbf{b})a_j\delta_{ik}
+ (2b^2-15a^2-2\,\mathbf{a}\cdot\mathbf{b})b_k\delta_{ij} 
\\
+(20b^2-5a^2-16\,\mathbf{a}\cdot\mathbf{b})a_i\delta_{jk}
+(20a^2-5b^2-16\,\mathbf{a}\cdot\mathbf{b})b_i\delta_{jk}
-(5a^2+b^2-26\,\mathbf{a}\cdot\mathbf{b})a_k\delta_{ij}
-(5b^2+a^2-26\,\mathbf{a}\cdot\mathbf{b})b_j\delta_{ik} 
\\
+4\,a_i\,a_j\,a_k+4\,b_i\,b_j\,b_k+12\,a_i\,a_j\,b_k+12\,b_i\,a_j\,b_k
-2\,a_i\,b_j\,a_k-2\,b_i\,b_j\,a_k-16\,a_i\,b_j\,b_k-16\,b_i\,a_j\,a_k\Bigr).
\end{multline}
Analogous calculations for the inertial range were performed by \cite{KZ-JoT}.

\subsection{Taking the Variational Derivative}
To take the variational derivative, we make a formal functional consequence from (\ref{1}):
\begin{equation}
B_q(\mathbf{r},t) = \varepsilon_{q i_1 j_1} \varepsilon_{j_1 l_1 k_1} \int\limits_{-\infty}^t\d_{i_1}
\bigl(v_{l_1}(\mathbf{r},\tau) B_{k_1}(\mathbf{r},\tau)\bigr)d\tau
 + \varkappa \,\d_{i_1}
\dd_{i_1} \int\limits_{-\infty}^t B_q(\mathbf{r},\tau)d\tau.
\end{equation}
The variational derivative of this expression is:
\begin{multline}
\frac{\delta B_q(\mathbf{r},t) }{\delta v_j(\mathbf{r}_1,t_1)} =
 \varepsilon_{q i_1 j_1}
\varepsilon_{j_1 l_1 k_1} \int\limits_{-\infty}^t\d_{i_1} \bigl(\delta_{j l_1}
\delta(\mathbf{r}-\mathbf{r}_1)\delta(\tau-t_1) B_{k_1}(\mathbf{r},\tau)\bigr)d\tau 
\\
+\varepsilon_{q i_1 j_1} \varepsilon_{j_1 l_1 k_1} \int\limits_{t_1}^t\d_{i_1}
\left(v_{l_1}(\mathbf{r},\tau)\frac{\delta B_{k_1}(\mathbf{r},\tau) }{\delta v_j(\mathbf{r}_1,t_1)}
\right)d\tau
+ \varkappa\, \d_{i_1} \d_{i_1} \int\limits_{t_1}^t\frac{\delta B_q(\mathbf{r},\tau) }
{\delta v_j(\mathbf{r}_1,t_1)}d\tau.
\end{multline}
The $\delta$-functions here are 'real', not regularized. The lower limits of the integrals in the
last two summands are changed, in accordance with the causality principle; after multiplying by
$\delta_{\epsilon}$ in the velocity correlators in (\ref{Fur-Nov-detailed}),  these lower limits
will make no difference. Taking the second derivative, we get:
\begin{multline}
\frac{\delta^2 B_q(\mathbf{r},t) }{\delta v_j(\mathbf{r}_1,t_1)\delta v_k(\mathbf{r}_2,t_2)}
 \stackrel{t_1, t_2\to t-0}{\simeq}
\varepsilon_{q i_1 j_1} \varepsilon_{j_1 j k_1}
\int\limits_{-\infty}^t d\tau_1 \,\d_{i_1} \left(\delta(\mathbf{r}-\mathbf{r}_1)\,\delta(\tau-t_1)\,
 \frac{\delta B_{k_1}(\mathbf{r},t)  }{\delta v_k(\mathbf{r}_2,t_2)}\right) 
\\
=\varepsilon_{q i_1 j_1} \varepsilon_{j_1 j k_1}\varepsilon_{k_1 i_2 j_2} \varepsilon_{j_2 k k_2}\,
\theta(t-t_1)\,\theta(t-t_2)\,
\times \d_{i_1} \bigl(\delta(\mathbf{r}-\mathbf{r}_1)\d_{i_2}
 (\delta(\mathbf{r}-\mathbf{r}_2) B_{k_2})\bigr).
\end{multline}
Here we omit the terms that become zero being multiplied by the regularized $\delta$-function.
Now we substitute these expressions into (\ref{Fur-Nov-detailed}). 
For the regularized delta functions we have
\begin{equation}
\int \limits_{-\infty}^{+\infty}  dt_1\,\int \limits_{-\infty}^t d\tau \,\delta_\eps(t-t_1)\,\delta(\tau-t_1)
=\int \limits_{-\infty}^{+\infty}  dt_1 \,\delta_\eps(t-t_1)\,\theta(t-t_1)
=\int \limits_{-\infty}^{t}  dt_1
\,\delta_\eps(t-t_1) =\frac{1}{2},
\end{equation}
\begin{equation}
\int \limits_{-\infty}^{t}  dt_1\,\int \limits_{-\infty}^{t}  dt_2
 \,\delta_\eps(t-t_1) \,\delta_\eps(t-t_2)=\frac{1}{4}
\int \limits_{-\infty}^{t}  dt_1\,\int \limits_{-\infty}^{t}  dt_2\,\delta_\eps(t-t_1) \,\delta_\eps(t_1-t_2)
=\int \limits_{0}^{+\infty}  dx\,\int \limits_{t_1-t}^{+\infty}  dy\,\delta_\eps(x) \,\delta_\eps(y)=\frac{3}{8}.
\end{equation}
These relations allow to take the time integrals in (\ref{Fur-Nov-detailed}). Thus, we arrive to
(\ref{FurNovikov_3}).
 Taking the space integrals, we get the closed equation for the pair correlator:
\begin{multline}\label{most-long}
\frac{\d}{\d t} \ls B_\alpha B'_\beta \rs  = 2\varkappa\, \d^{\rho}_m \d^{\rho}_m
\ls B_\alpha B'_\beta \rs  
\\
+\frac 12
 \bigl(\varepsilon_{\alpha mn} \varepsilon_{n p q} \delta_{\beta r}+
 \varepsilon_{\beta mn} \varepsilon_{n p q} \delta_{\alpha r}\bigr)
  \varepsilon_{j_1 j k_1}  \d^{\rho}_m\d^{\rho}_{i_1}
\left(\varepsilon_{q i_1 j_1} D_{jp}(0)
 \ls{B_{k_1} B'_r}\rs -\varepsilon_{r i_1 j_1}
D_{jp}(\brho)
 \ls{B_q B'_{k_1}}\rs   \right) 
 \\
+\frac 16
\bigl(\varepsilon_{\alpha mn} \varepsilon_{n p q} \delta_{\beta r}
+ \varepsilon_{\beta mn} \varepsilon_{n p q}   \delta_{\alpha r}\bigr)
 \varepsilon_{j_1 j k_1}\varepsilon_{j_2 k k_2}
\d^{\rho}_m\Bigl(2\varepsilon_{q i_1 j_1} \varepsilon_{r i_2 j_2}
\d^{\rho}_{i_2}
\bigl(
\d^{(2)}_{i_1}   F_{kpj}(\brho,\brho)
\ls{  B_{k_1}  B'_{k_2}}\rs 
\\
+ F_{kpj}(\brho,\brho)
\d^{\rho}_{i_1}    \ls{  B_{k_1}  B'_{k_2}}  \rs   \bigr)
+\varepsilon_{r i_1 j_1}\varepsilon_{k_1 i_2 j_2}
\d^{\rho}_{i_1} \bigl(
\d^{(2)}_{i_2}   F_{pjk}(\brho,\brho)
  \ls{B_{q}B'_{k_2}}\rs
+F_{pjk}(\brho,\brho)
 \d^{\rho}_{i_2}   \ls{B_{q}B'_{k_2}}\rs  \bigr)\Bigr),
\end{multline}
where $\brho = \br'-\br$ and $\d^{(2)}_n $  denotes the derivative over the second argument.

Making use of (\ref{DcherezP}), we express all the coefficients in the (\ref{most-long}) by means
of the function $P(r)$ and the longitudinal velocity structure function.  After cumbersome symbol
math-assisted calculations, we
 arrive to the
generalized Kazantsev equation: (\ref{dQdt_v3-1}), (\ref{27-5}), (\ref{deltaL-simple}) for the
Batchelor regime and (\ref{dQdt_v3-1}), (\ref{Lgauss-nonlin}), (\ref{deltaL-tilde}) for the
nonlinear velocity field.

\section{Solutions of the generalized Kazantsev equation}

\subsection{Kazantsev-Kraichnan Model, Linear Velocity Field}
Consider first the simplest Kazantsev equation for the Kazantsev-Kraichnan model (\ref{Kaz-eq})
in the Batchelor regime, i.e., with $\sigma = (D/3) r^2$ (which corresponds to $r_{\nu}\to \infty$).
The substitution of the ansatz
$$
G (t,r) = e^{\gamma t} G(r)
$$
 reduces this equation to the ordinary differential equation
$$
\gamma G =\frac 23 D \left( r^2 G'' + 6 r G' +10 G \right) +2\varkappa \left( G''+ \frac 4r G' \right).
$$
We proceed to the dimensionless variables $x=r/r_d$; with account of $\varkappa = D r_d^2$, we get
\begin{equation} \label{AppC-1}
 \left( x^2 +3 \right) G'' + \left(6 x + \frac {12}x  \right) G'  + \left( 10-\frac 32 \Gamma \right) G =0,
\end{equation}
where
$$
\Gamma = \gamma/D.
$$
The two exact solutions of this equation can be written by means of hypergeomertic functions.
Here we restrict our consideration to the analysis of their asymptotic behavior, to get the experience
necessary to generalize the solutions to the cases of nonzero $F$ and/or finite $r_{\nu}$.

One of two independent solutions diverges as $x \to 0$, so we are interested in the other one.
It satisfies the boundary condition $G'(0)=0$.

\begin{figure}[h]   
\centering
\psfrag{g}{$\Gamma$} \psfrag{l1}{Im $\alpha_+$} \psfrag{l2}{Re $\alpha_+$}
\psfrag{0}{$0$}\psfrag{a1}{$5/2$}\psfrag{a2}{$20/3$}\psfrag{a3}{$-\frac{5}{2}$}
\psfrag{a4}{$\frac{\sqrt{15}}{2}$}
\includegraphics[width=8cm]{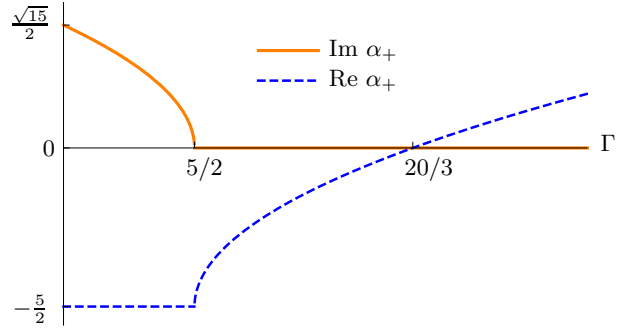}
\caption{Real (orange) and imaginary (blue) parts of $\alpha_+$ for different $\Gamma>0$.  }
\label{fig6}
\end{figure}

As $x \to \infty$, the equation (\ref{AppC-1}) is simplified to a homogenous differential equation;
its characteristic equation is
$$
\alpha (\alpha-1) + 6 \alpha + 10 - \frac 32 \Gamma = 0 \ ,
$$
and the solution is
$$
G^{(\Gamma)}(x) \simeq_{x\to \infty}   A_+ x^{\alpha_+} + A_{-} x^{\alpha_-},
$$
where
$$
\alpha_{\pm}=-\frac52\pm\frac{\sqrt{3}}2\sqrt{2\Gamma-5}.
$$
The real and imaginary parts of $\alpha_+(\Gamma)$ are illustrated in Figure~\ref{fig6}. One can see that
 for all $\Gamma <  5/2$, $G^{(\Gamma)}(x)$  decreases as $x\to \infty$ with
 the same rate and oscillates, while for $\Gamma >5/2$ it decreases slower
  (and for $\Gamma>20/3 $ even grows), without oscillations.

Now, let us return to the evolution equation (\ref{Kaz-eq}).
The equation (\ref{AppC-1}) can be reduced to the Sturm-Liouville equation (by the change of variables
 $y=arsinh (x/\sqrt{3})$, $q(y)=(3+x^2)^{1/4} x^2 G(x)$).
 So, an arbitrary initial perturbation $G_0 (x)$
can be decomposed into a sum (integral) of the eigenfunctions $G^{(\Gamma)}(x)$.
The time dependence of each summand is exponential
with its own rate, $G^{(\Gamma)}(t,x) \propto e^{\gamma t}$, so the term with the biggest
 $\gamma$ would survive after large time. Now, if the initial function is localized (more precisely,
  decreases not faster than $x^{-5/2}$ as $x \to \infty$),
then all the terms in the decomposition must oscillate or decrease at $x\to \infty$
faster than $G_0(x)$. Since ${\rm Im} \alpha_+ (\Gamma \ge 5/2) =0$ and ${\rm Re} \alpha_+ (\Gamma \ge 5/2) >-5/2$,
we get the upper boundary of the spectrum:
$$
\Gamma_{max} = \frac 52 \ , \quad G(x,t) \propto e^{\Gamma_{max} D t}.
$$

\subsection{Non-Zero Time Asymmetry}

Now we take account of $F\ne 0$, and
apply the same ideas to analyze the equation (\ref{dQdt_v3-1}), (\ref{deltaL-simple}).
The equation for the eigenvalues is
\begin{equation} \label{AppC-asymEq}
\gamma G =\frac 23 D \left( r^2 G'' + 6 r G' +10 G \right) +2\varkappa \left( G''+ \frac 4r G' \right)
+\frac 19 F \left( 2r^3 G'''+21 r^2 G''+ 14r G' -70 G \right).
\end{equation}
In the dimensionless variables, this is equivalent to
\begin{equation} \label{AppC-2}
 \left( x^2 +3 \right) G'' + \left(6 x + \frac {12}x  \right) G'  + \left( 10-\frac 32 \Gamma
\right) G \\
+\frac 16 f \left( 2x^3 G'''+21 x^2 G''+ 14x G' -70 G \right)  =0,
\end{equation}
where $f=F/D$ is the new small parameter. Again, we consider the asymptote $x\to \infty$ and get the
homogenous differential equation with the characteristic polynom:
\begin{equation} \label{char-pol}
\frac{f}{3}\beta(\beta-1)(\beta-2)+\left(1+\frac72f\right)\beta(\beta-1)
+\left(6+\frac73f\right)\beta+\left(10-\frac{35}{3}f-\frac32\Gamma\right)=0.
\end{equation}
This equation has three solutions, two of them are close to $\alpha_{\pm}$, 
and the third solution is real (negative) and very big:
\begin{align}
G^{(\Gamma)}&\simeq B_+\, \xi^{\beta_+}+B_- \,\xi^{\beta_-}+B_f \,\xi^{\beta_f}, \qquad \xi\to\infty, \\
&\beta_f\propto -1/f, \\
&\beta_{\pm}-\alpha_\pm\propto f. \label{E:QInf_Fne0}
\end{align}

%
The third summand is an artefact of the model; it corresponds to the non-physical solution and must
 be excluded by setting $B_f=0$.

 The internal limit $x \to 0$ gives also three solutions, only one of them is convergent. We assume that
 there exist  modes that converge at both limits $x \to 0$ and $x \to \infty$. Unlike the Gaussian
 (Kazantsev-Kraichnan) case, this is not guaranteed. This supposition is to be proved by numerical
 simulations.  This is done in Section V. Also, the completeness of this set of functions is not evident; however, we suppose
that the asymptotic time dependence of arbitrary solution is exponential, as it was in the case of
the second
order equation, so this arbitrary solution can be presented as a sum of eigenfunctions; we are again
interested in the eigenfunction with the fastest increment.

But as soon as we suppose the existence of the spectrum, we can derive the upper boundary of the
eigenvalue in the decomposition of an arbitrary initial distribution $G_0(x)$ based on the asymptotic
behavior of the eigenfunctions.
The eigenfunctions presented in the decomposition can either decrease faster than $G_0(x)$ of oscillate
as $x \to \infty$.
From the characteristic polynom (\ref{char-pol}) we find that the condition of oscillations
${\rm Im} \beta \ne 0$ holds for
\begin{equation}
\Gamma<\Gamma_{{max}}=\frac{8+126f^2-\left(4+27f^2\right)^{3/2}}{18f^2}.
\end{equation}
This coincides with (\ref{E:gamma_full}).
In the limit $f\ll 1$, we arrive to the Equation (\ref{gammaFbatch}).

\subsection{Finite Prandtl Numbers}
To evaluate the effect of the nonlinearity of velocity field,  we consider the function $\sigma(r)$
truncated in accordance with (\ref{sigma-ranges-late}). For this model, the Kazantsev equation
is step-wise: it coincides with (\ref{AppC-asymEq}) for $r<r_{\nu}$, while for $r>r_{\nu}$ it becomes
\begin{equation}
\gamma G = \frac 23 D (G''+\frac 4r G') (r_{\nu}^2+2 r_d^2)  \ , r>r_{\nu}.
\end{equation}
So, one has to match the solution of (\ref{AppC-asymEq}) with the descending solution of this equation.
The coincidence of $G$ and $G'$ determines the spectrum of $\gamma$. But the biggest
$\gamma_{max}^{(\mu)}$ is
still restricted by the condition ${\rm Im} \beta \ne 0$.

To estimate the correction to $\gamma$ produced by large but finite $r_{\nu}$, we note that
 the solution of (\ref{AppC-asymEq}) oscillates as a function of $r$,
 and the phase of the oscillations is determined by the imaginary part of $\beta_+$:
\begin{equation}
 G(r) \propto r^\beta_+ \propto  e^{i {\rm Im} \beta_+ \ln \mu}.
\end{equation}
Matching the amplitudes of both branches of the solution at $r=r_{\nu}$ is provided by a multiplier;
to match the derivatives, one has to choose the phase. However, half a period of oscillations is enough
to ensure any phase that is needed. So, the correction to $\gamma_{max}$ produced  by finite $r_{nu}$ at
any rate leaves it within the range of $\gamma$ in which the phase of $G(r_{\nu})$ changes by $\pi$:
\begin{equation}
{\rm Im} \beta_+ (\Gamma_{max}^{(\mu)}) \ln \mu = \pi.
\end{equation}
The bigger $\mu$, the closer is $\Gamma_{max}^{(\mu)}$ to the 'Batchelor' value  $\Gamma_{max}$.

\end{document}